\documentclass{pasj00}
\draft

\begin{document}
\SetRunningHead{Y. Takeda et al.}{On the Na vs. Fe Correlation in Late B-Type Stars}
\Received{2013/08/20}
\Accepted{2013/10/08}

\title{On the Sodium versus Iron Correlation \\
in Late B-Type Stars
\thanks{Based on data collected at Okayama Astrophysical Observatory (NAOJ, Japan).}
}

%

\author{
Yoichi \textsc{Takeda,}
Satoshi \textsc{Kawanomoto,} and
Naoko \textsc{Ohishi}
}
\affil{National Astronomical Observatory, 2-21-1 Osawa, 
Mitaka, Tokyo 181-8588}
\email{takeda.yoichi@nao.ac.jp, kawanomoto.satoshi@nao.ac.jp, naoko.ohishi@nao.ac.jp}

%

\KeyWords{
physical processes: diffusion --- stars: abundances  \\
--- stars: atmospheres --- stars: chemically peculiar --- stars: early-type} 

\maketitle

\begin{abstract}
With an aim to study whether the close correlation between [Na/H] and [Fe/H] 
recently found in A-type stars further persists in the regime of B-type stars, 
the abundances of Na were determined for 30 selected sharp-lined late B-type 
stars (10000~K~$\ltsim T_{\rm eff} \ltsim 14000$~K) from the 
Na~{\sc i} 5890/5896 doublet. These Na abundances were then compared 
with the O and Fe abundances (derived from the O~{\sc i} 
6156--8 and Fe~{\sc ii} 6147/6149 lines) showing anti-correlated peculiarities.
It turned out that, unlike the case of A-type stars, [Na/H] is roughly constant at
a slightly subsolar level ([Na/H]~$\sim -0.2 (\pm 0.2)$ without any significant 
correlation with [Fe/H] which shows considerable dispersion ranging from 
$\sim -0.6$ to $\sim +1.0$. This may serve as an important observational 
constraint for understanding the abundance peculiarities along with
the physical mechanism of atomic diffusion in upper main-sequence 
stars of late A through late B-type including Am and HgMn stars.
\end{abstract}

%


\section{Introduction}

Abundance studies of sodium (Na) for early-type stars in the upper main 
sequence have been insufficient, despite the astrophysical importance
of this element. Actually, until recently, investigations of Na abundances have 
long been limited up to late A stars ($T_{\rm eff} \ltsim 8500$~K), and those 
of higher $T_{\rm eff}$ from early A through B remained essentially untouched.

This is presumably due to its elemental characteristics; i.e., an alkali element
with only one valence electron weakly bound and thus its ionization potential
is fairly low ($\sim$~5.1~eV), which means that almost all Na atoms are ionized 
and only a very small fraction remain neutral in the atmospheres of early-type stars.
Accordingly. several subordinate Na~{\sc i} lines of low excitation 
($\chi_{\rm low} \sim 2$~eV) frequently used for abundance determinations in
late-type FGK stars are too weak to be measurable at higher $T_{\rm eff}$, while
the strong resonance lines (Na~{\sc i} 5890/5896, so called D lines) tend to 
be avoided because they are not easy to handle (e.g., non-LTE effect, contamination 
of interstellar lines, strong parameter-sensitivity, etc.). 

Nevertheless, an extensive Na abundance study based on non-LTE 
analysis of strong Na~{\sc i} 5890/5896 lines was challenged by Takeda et al. 
(2009; hereinafter referred to as Paper~I) for 122 unbiased sample of 
A-type stars (7000~K~$\ltsim T_{\rm eff} \ltsim 10000$~K) having a wide range of 
rotational velocities. 
Unfortunately, the resulting abundances turned out systematically 
subsolar ($-1 \ltsim$~[Na/H]~$\ltsim 0$) and unreliable, which was eventually
interpreted as due to the considerable sensitivity of Na abundances (derived
from such strongly saturated Na~{\sc i} D lines) to a choice of microturbulence 
along with its inevitable uncertainty due to depth-dependence (see subsection 5.3 
in Paper~I and the references therein).

Having learned a lessen for this problem, Takeda et al. (2012; hereinafter 
referred to as Paper~II) then determined the abundances of Na (along with other 
alkalis such as Li and K) based on weaker subordinate Na~{\sc i} 5682/5688 
lines (less sensitive to microturbulence) for 24 selected sharp-lined A-type 
stars ($v_{\rm e}\sin i \ltsim 50$~km~s$^{-1}$, 
7000~K~$\ltsim T_{\rm eff} \ltsim 10000$~K). 
Interestingly, the resulting Na abundances were found to tightly scale with 
those of Fe (cf. figure 11a in Paper~II), despite that Fe abundances 
were quite diversified in those 24 sample stars including Am stars
(generally showing Fe enrichment). Actually, this tendency is considered to 
hold for A-type stars in general, since figure 7b in Paper~I definitely shows
evidence of such a positive correlation (even if [Na/H] values themselves
may suffer systematic errors).

This correlation between Na and Fe has a significant implication on the understanding
of the physical mechanism for producing the abundance anomalies of Am stars 
(a group of A-type stars generally of slow rotation; see, e.g., Preston 1974
for a review), where heavier elements such as iron-group are overabundant 
while light elements such as CNO tend to be deficient. 
It is generally believed that some kind of segregation
process (atomic diffusion) in the stellar envelope (depending on the balance 
between the downward gravitational settlement and upward radiative acceleration)
is responsible for these peculiarities. However, since none of the existing 
theoretical calculations for Am stars (e.g., Richer et al. 2000 or 
Talon et al. 2006) has predicted that Fe and Na behave in the same way,
this observational facts may require that the theory should be improved. 
How come these two elements (where atomic level configurations considerably
differ from each other) show similar anomalies? 

Here, we should recall that another group of apparently 
non-magnetic chemically peculiar stars of somewhat higher $T_{\rm eff}$ 
exists at the spectral type of late B, which are namely HgMn stars 
(cf. the review by Preston 1974) sharing rather similar characteristics 
to those of Am stars; 
such as an underabundance of O coupled with an overabundance of Fe 
(e.g., Takeda et al. 1999). Actually, the possible connection between 
these two groups has been occasionally argued (e.g., Adelman, Adelman, \& 
Pintado 2003).
If so, it is worthwhile to check whether or not such a close
correlation between Na and Fe is seen also in sharp-lined late B-type 
stars including HgMn stars. 

As already mentioned, very few people have tried determinations of 
Na abundances for B-type stars so far, except for the recent LTE 
studies of Fossati et al. (2009) for two sharp-lined late-B 
stars (21~Peg and $\pi$~Cet). Given this situation, we decided to 
carry out a new spectroscopic study for the abundances of O, Fe, 
and Na (where O and Fe abundances are indicators for HgMn peculiarities) 
for selected 30 sharp-lined late B-type stars, with an aim to give 
a clear answer to this question (i.e., whether or not Na correlates 
with Fe also in late B-type stars). This time, since subordinate lines 
such as Na~{\sc i} 5682/5688 are too weak to be measurable, 
we have no other way than to invoke the resonance lines of 
Na~{\sc i} 5890/5896, which are fortunately moderately weak and 
not strongly saturated at this range of $T_{\rm eff}$; 
thus, we do not suffer such problems as encountered in Paper~I.
The purpose of this paper is to report the outcome of this 
investigation.

\section{Observational Data}

As the targets of this study, we selected 30 late B-type 
stars (including a few A0 stars) mostly with spectral classes 
of B6--B9.5 and luminosity classes of III--V (cf. table 1),
which are comparatively sharp-lined for early-type stars 
($v_{\rm e}\sin i \ltsim 60$~km~s$^{-1}$). 

For 9 stars among these, we could avail ourselves of old spectra
covering a wavelength range of 5600--6800~$\rm\AA$ 
with a resolving power of $R \sim 70000$, which were 
obtained in 2006 October by using HIDES spectrograph at
Okayama Astrophysical Observatory (OAO) to determine 
the O and Ne abundances of B-type stars. See Takeda et al. 
(2010) for more details.

Regarding the other 21 stars, we carried out new observations 
on 2012 May 26--31 again by using OAO/HIDES, where the same slit 
width (200~$\mu$m) as in 2006 October observations was adopted 
to get spectra of $R\sim 70000$ covering the
wavelength range of 5100--8800~$\rm\AA$.\footnote{Note that at the 
time of 2006 observation, the wavelength coverage of HIDES was only 
$\sim$~1200~$\rm\AA$ because only one CCD was available. At present, HIDES has 
three mosaicked 4K$\times$2K CCDs, accomplishing the whole wavelength 
coverage of $\sim 3700$~$\rm\AA$ in the mode of red cross-disperser.} 
The reduction of these new 2012 spectra was done in the same manner as 
in  2006 data, and S/N ratios typically on the order of $\sim 200$ 
were attained.
For both of the 2006 and 2012 data, telluric lines existing in the region 
of Na~{\sc i} 5890/5896 lines were removed by dividing the spectra
by that of a rapid rotator as done in Paper~I. 

These 30 program stars are plotted on the $\log L$ vs. $\log T_{\rm eff}$ 
diagram (theoretical HR diagram) in figure 1, where Girardi et al.'s (2000)
theoretical evolutionary tracks corresponding to different stellar 
masses are also depicted. We can see from this figure that the masses 
of our sample stars are in the range between $\sim 2.5 M_{\odot}$ and 
$\sim 5 M_{\odot}$.

\section{Atmospheric Parameters}

In order to maintain consistency with our recent studies (Paper~I, 
Paper~II, Takeda et al. 2010), the effective temperature ($T_{\rm eff}$) 
and the surface gravity ($\log g$) of each program star were 
determined from the colors of Str\"{o}mgren's $uvby\beta$ photometric 
system with the help of Napiwotzki, Sc\"{o}nberner, and Wenske's (1993) 
{\tt uvbybetanew} program\footnote{
$\langle$http://www.astro.le.ac.uk/\~{}rn38/uvbybeta.html$\rangle$.}, 
where the observational data of $b-y$, $c_{1}$, $m_{1}$, and $\beta$ 
were taken from Hauck and Mermilliod (1998) via the SIMBAD database.
The resulting $T_{\rm eff}$ and $\log g$ are summarized in table 1.
Their typical errors may be estimated as $\sim 3\%$ in 
$T_{\rm eff}$ and $\sim 0.2$~dex in $\log g$ for the present
case of late B stars, according to Napiwotzki et al. 
(1993; cf. their section 5).\footnote{
As a check, we also tried deriving $T_{\rm eff}$ and $\log g$ by using 
Moon's (1985) {\tt UVBYBETA/TEFFLOGG} program based on 
Moon and Dworetsky's (1985) calibration, which may be better for stars 
of the relevant spectral type according to Kaiser (2006). We then found
that the resulting values of $T_{\rm eff}$ are slightly (but systematically) 
higher by $\sim 200$~K ($\langle \Delta T_{\rm eff} \rangle = 208 \pm 124$~K) 
while those of $\log g$ are practically the same
($\langle \Delta \log g \rangle = -0.03 \pm 0.05$~dex) ,
compared to our adopted parameters based on the Napiwotzki et al.'s (1993) 
{\tt uvbybetanew} program. Therefore, we should bear in mind the 
possibility that our $T_{\rm eff}$ given in table 1 might be
slightly underestimated by $\sim 200$~K on the average, though 
this is still within the considered uncertainty ($\sim \pm 3\%$).}

The model atmosphere for each star was then constructed
by two-dimensionally interpolating Kurucz's (1993) ATLAS9 
model grid in terms of $T_{\rm eff}$ and $\log g$, where
the solar-metallicity models were exclusively used as
in our previous studies. 

Regarding the microturbulent velocity ($\xi$), we assume 
$\xi = 1$~km~s$^{-1}$ with uncertainties of $\pm 1$~km~s$^{-1}$ 
in this study,\footnote{In contrast, Takeda et al. (2010) assumed 
$\xi = 3$~km~s$^{-1}$ with uncertainties of $\pm 2$~km~s$^{-1}$ 
based on the results of Lyubimkov, Rostopchin, and Lambert (2004), 
since B-type main-sequence stars in a much wider $T_{\rm eff}$ range
(10000~K~$\ltsim T_{\rm eff} \ltsim 23000$~K, corresponding to 
masses of 4--11~$M_{\odot}$) 
were involved in that investigation.} which we regard as most 
reasonable considering the published $\xi$ results for late B-type 
stars (10000~K~$\ltsim T_{\rm eff} \ltsim 15000$~K) typically 
ranging between $\sim 0$ and $\sim 2$~km~s$^{-1}$ (see, e.g., 
table 1 of Sadakane 1990).

\section{Abundance Determinations}

The determination procedures of elemental abundances and related
quantities (e.g., non-LTE correction, uncertainties due to
ambiguities of atmospheric parameters) are essentially the same 
as adopted in Takeda et al. (1999), Papers I and II, and 
Takeda et al. (2010), which consist of two consecutive steps. 

\subsection{Synthetic spectrum fitting}

The first step is to find the solutions for the abundances of 
relevant elements, projected rotational velocity ($v_{\rm e}\sin i$), 
and radial velocity ($V_{\rm rad}$) such as those accomplishing the
best fit (minimizing $O-C$ residuals) between theoretical and 
observed spectra, while applying the automatic fitting algorithm 
(Takeda 1995). Two wavelength regions were selected for this purpose:
[1] 6145--6160~$\rm\AA$ region comprising O~{\sc i} 6156--8 and 
Fe~{\sc ii} 6147/6149 lines, which is to determine the abundances
of oxygen and iron, and [2] 5888--5898~$\rm\AA$ region comprising
Na~{\sc i} 5890/5896 lines to establish the abundance of Na.
The adopted atomic data of important lines are presented in table 2.
How the theoretical spectrum for the converged solutions fits well 
with the observed spectrum is displayed in figures 2 (region [1]) 
and 3 (region [2]). The $v_{\rm e}\sin i$ values\footnote{It should be 
kept in mind that we assumed only the rotational broadening (with the 
limb-darkening coefficient of $\epsilon = 0.5$) as the macrobroadening 
function to be convolved with the intrinsic theoretical line profiles.
Accordingly, $v_{\rm e}\sin i$ values for very sharp-line cases 
(e.g., $v_{\rm e} \sin i \ltsim$~5--6~km~s$^{-1}$) may be somewhat
overestimated because the effects of the instrumental broadening 
or of the macroturbulence may not be negligible compared to 
$v_{\rm e}\sin i$.} resulting from region [1] fitting are 
presented in table 1.

Note that, in the evaluation of $O-C$ residuals, we had to occasionally 
mask some regions showing features irrelevant to stellar spectra; 
such as spurious spectrum defect (figure 2) or interstellar absorption 
components of Na~{\sc i} 5890/5896 lines (figure 3), as 
highlighted in green in both figures. Also, in the case of 
5888--5898~$\rm\AA$ fitting of Na~{\sc i} lines, $v_{\rm e}\sin i$ 
values were often fixed at the values derived from 6145--6160~$\rm\AA$
fitting, since only incomplete Na~{\sc i} D line profiles
are available in most cases.

\subsection{Analysis on inversely evaluated $W_{\lambda}$}

As the second step,  with the help of Kurucz's (1993) WIDTH9 program 
(which had been considerably modified in various respects; e.g., 
inclusion of non-LTE effects, treatment of total equivalent width for 
multi-component lines; etc.), we computed the equivalent widths of 
the representative lines ``inversely'' from the abundance solutions
(resulting from spectrum synthesis) along with the adopted atmospheric 
model/parameters; i.e., $W_{6147}^{\rm Fe}$ (for Fe~{\sc ii} 6147), 
$W_{6156-8}^{\rm O}$ (for O~{\sc i} 6156--8 merged triplet),
and $W_{5890}^{\rm Na}$ (for Na~{\sc i} 5890).

Since the non-LTE effect was explicitly taken into consideration 
for O and Na as done in Papers I, II, and Takeda et al. (1999, 2010), 
$A_{\rm N}$ (NLTE abundance) and $A_{\rm L}$ (LTE abundance) were
computed from $W_{6156-8}^{\rm O}$ and $W_{5890}^{\rm Na}$, from 
which the NLTE correction $\Delta (\equiv A_{\rm N} - A_{\rm L})$
was further derived. The resulting NLTE abundances and corrections
for O and Na, as well as the LTE abundances for Fe, are given in table 1.
Although only the results for the stronger Na~{\sc i} 5890 line of
the doublet are given in table 1, those for the weaker Na~{\sc i} 5896
are well represented by the following formula:
\begin{equation}
W_{5896} = 6.74\times10^{-1} + 4.68\times10^{-1} W_{5890} 
   + 2.96\times10^{-3} W_{5890}^{2}
\end{equation}
where $W_{5896}$ and $W_{5890}$ are measured in m$\rm\AA$, and
\begin{equation}
\Delta_{5896} = -0.268 - 0.254 \Delta_{5890} 
   -0.946 \Delta_{5890}^{2}.
\end{equation}
As expected, inequality relations of $W_{5896} < W_{5890}$ and 
$|\Delta_{5896}| < |\Delta_{5890}|$ generally hold.

We then estimated six kinds of abundance variations
($\delta_{T+}$, $\delta_{T-}$, $\delta_{g+}$, $\delta_{g-}$, 
$\delta_{\xi+}$, and $\delta_{\xi-}$) for $A^{\rm O}$, $A^{\rm Na}$,
and $A^{\rm Fe}$ by repeating the analysis on the $W$ values while 
perturbing the standard atmospheric parameters interchangeably by 
$\pm 3\%$ in $T_{\rm eff}$, $\pm 0.2$~dex in $\log g$, 
and $\pm 1$~km~s$^{-1}$ in $\xi$ (which are the typical 
ambiguities of the parameters we adopted; cf. section 3). 
Finally, the root-sum-square of these perturbations,
$\delta_{Tg\xi} \equiv (\delta_{T}^{2} + \delta_{g}^{2} + \delta_{\xi}^{2})^{1/2}$, 
were regarded as abundance uncertainties (due to combined errors in 
$T_{\rm eff}$, $\log g$, and $\xi$), 
where $\delta_{T}$, $\delta_{g}$, and $\delta_{\xi}$ are defined as
$\delta_{T} \equiv (|\delta_{T+}| + |\delta_{T-}|)/2$, 
$\delta_{g} \equiv (|\delta_{g+}| + |\delta_{g-}|)/2$, 
and $\delta_{\xi} \equiv (|\delta_{\xi+}| + |\delta_{\xi-}|)/2$,
respectively. 

Figures 4(O), 5(Na), and 6(Fe) graphically show the resulting abundances,
equivalent widths, non-LTE corrections (or $v_{\rm e}\sin i$ in figure 6), 
and abundance variations in response to parameter changes, 
as functions of $T_{\rm eff}$. We can see from these figures that, 
while the extents of non-LTE corrections for O~{\sc i} 6156--8 are 
not very significant (typically $\sim$~0.1--0.2~dex), those for 
Na~{\sc i}~5890 are significantly large and non-negligible 
($\sim$~0.3--0.5~dex).
Besides, the abundance errors due to parameter uncertainties are
typically $\sim$~0.1--0.2~dex for Na and Fe (though errors in Na 
may grow up to $\ltsim$~0.3~dex for near-A0 stars with 
$T_{\rm eff} \ltsim 11000$~K), while those for O are not important
($\ltsim 0.1$~dex). 

\section{Discussion}

\subsection{Trend of O/Na/Fe Abundances with $v_{\rm e} \sin i$}

In order to discuss the results derived in section 4,
the mutual relations between [O/H], [Na/H], and [Fe/H] (the differential 
abundances relative to the Sun) as well as their behaviors with respect 
to $v_{\rm e}\sin i$ are displayed in figure 7, where stars with 
$T_{\rm eff} < 11000$~K and $T_{\rm eff} > 11000$~K are discriminated
in different symbols.

We can barely recognize the weak $v_{\rm e}\sin i$-dependence
in [O/H] (figure 7d) and [Fe/H] (figure 7f) (i.e., O tends to be 
deficient while Fe to be overabundant with a decrease in $v_{\rm e}\sin i$).
Besides, figure 7a manifestly shows that [O/H] and [Fe/H] are
anti-correlated with each other in the sense that [O/H] gets
decreased from $\sim 0.0$ to $\sim -0.7$ as [Fe/H] increases
from $\sim 0.0$ to $\sim +1.0$. 
Since these tendencies are qualitatively the same as seen in sharp-lined 
A-type stars showing Am phenomenon (Takeda \& Sadakane 1997, Takeda et al. 
2008, Paper~I), we may state that O and Fe abundances of these sharp-lined 
late B-type stars (10000~K~$\ltsim T_{\rm eff} \ltsim 14000$~K) can be 
regarded as peculiarity indicators (weakly dependent upon $v_{\rm e}\sin i$, 
behaving in an opposite way to each other) just as has been reported for 
A-type stars (7000~K~$\ltsim T_{\rm eff} \ltsim 10000$~K). Note that
this consequence is consistent with the results of Takeda et al. (1999;
cf. figures 10 and 11 therein)

Now, we are almost ready to answer the question which originally 
motivated this investigation: ``Does [Na/H] scale with [Fe/H] in 
accordance with the extent of peculiarity in these late B stars,
just as seen in A stars?'' Our answer is negative according to figure 7b,  
which reveals that [Na/H] values are roughly homogeneous distributing 
at a mildly subsolar level between $\sim -0.5$ and $\sim 0.0$ 
(the average is $\langle$[Na/H]$\rangle = -0.19$ with $\sigma = 0.16$, 
where the outlier value of [Na/H] = +0.59 for HD~143807\footnote{
This star ($\iota$~CrB), being very sharp-lined with almost the lowest 
$v_{\rm e} \sin i$ (4~km~s$^{-1}$), belongs to the low $T_{\rm eff}$ group of near 
A0 (10828~K). Reflecting its conspicuously strong and saturated $W_{5890}$ 
(116~m$\rm\AA$), the Na abundance of this star is especially sensitive to
errors in $T_{\rm eff}$ as well as $\xi$, as manifestly seen in
figures 5d and 5f. Accordingly, we have to keep in mind that the Na abundance
derived for this star is less reliable compared to other stars, which might 
explain (at least partly) this outlier nature.} 
is excluded) without any significant dependence on [Fe/H] 
(showing a large dispersion from $\sim -0.6$ to $\sim +1.0$).
Yet, if we confine to stars with 10000~K~$ \ltsim T_{\rm eff} \ltsim 11000$~K 
(red circles in figure 7b), we still see a sign of weak positive correlation
between [Na/H] and [Fe/H], presumably because they tend to share the 
characteristics of A0-type stars. However, as $T_{\rm eff}$ becomes higher,
this trend eventually disappears at $T_{\rm eff} \gtsim 11000$~K (blue
triangles) where [Na/H] tends to be stabilized with only a small dispersion 
around [Na/H]~$\sim -0.2\; (\pm 0.2)$. Hence, we may state that whether and 
how Na correlate with Fe in upper main-sequence stars depends 
significantly on $T_{\rm eff}$. 

\subsection{$T_{eff}$-Dependence of Na vs. Fe Correlation}

Let us investigate this issue in a more quantitative manner. 
Combining the present results with those of our previous studies 
(Takeda 2007; Paper~II), we display in figure 8 the [Na/H] vs. [Fe/H] 
relations for four stellar groups of different $T_{\rm eff}$ ranges
(FGK-type stars, A-type stars, late B-type stars of near A0, and 
late B-type stars), and the statistical quantities showing 
the nature of correlation between these two elements were also 
computed for each group as given in table 3.
We can recognize from this table (and intuitively confirmed
from figure 8) that the slope ($m$) of linear-regression 
relation ([Na/H] $\sim m$ [Fe/H] +$b$) progressively decreases while 
the dispersion increases as $T_{\rm eff}$ becomes higher, which is 
an interesting tendency.
It should be remarked, however, that the characteristics
of FGK stars (first group) and those of A--B stars (last three groups) 
are intrinsically different from each other.

Regarding the former late-type stars in the Galactic disk 
($\sim$~1--1.5~$M_{\odot}$) which have ages of $\sim 10^{9}$--$10^{10}$~yr, 
the reason for the spread of $\sim 1$~dex in [Fe/H] as well as [Na/H] 
is due to the chemical evolution in our Galaxy. 
And the scaling relation of [Na/H]~$\sim$~[Fe/H] may be interpreted by 
the fact that the synthesis of Na (odd-Z neutron-rich element) is 
sensitive to neutron excess and thus considered to be metallicity-dependent.
Actually, [Na/Fe]$\sim 0$ almost holds for disk stars of 
$-1 \ltsim $~[Fe/H]~$\ltsim 0$, though a slight upturn is observed
at the metal-rich end ($0 \ltsim $~[Fe/H]~$\ltsim 0.5$) (see, e.g.,
Takeda et al. 2003).

In contrast, early-type stars with masses of 
$\sim$~1.5--5~$M_{\odot}$ are considerably younger 
($\ltsim 10^{8}$--$10^{9}$~yr), which are regarded to
have formed from gas of nearly the same chemical composition. 
(see, e.g., the near-homogeneity of O and Ne 
abundances in B-type stars derived by Takeda et al. 2010).
Accordingly, the diversified Na as well as Fe abundances
in these sharp-lined A and late-B stars 
($T_{\rm eff} \sim$~7000--14000~K) must be due to ``abundance 
peculiarities'' which were acquired during their main-sequence
life time presumably due to chemical segregation processes 
operating in the stable atmosphere or envelope. 
The significance of the result we found is, therefore, that
it can provide with important observational characteristics 
of chemical peculiarities in A and late-B stars to be explained
theoretically, e.g., by atomic diffusion calculations.

\subsection{Conclusion}

Regarding our primary aim of this investigation 
(i.e., to check whether or not the trend of Na vs. Fe 
correlation seen A-type stars persists into the regime 
of late B-type stars), our conclusion is summarized as follows:\\ 
--- (1) In the regime of A-type stars 
(7000~K~$\ltsim T_{\rm eff} \ltsim 10000$~K), 
the mechanism causing chemical peculiarities (mainly Am 
anomalies in this case, characterized by overabundance of Fe
and underabundance of O) acts nearly equally on both Na and Fe,
so that a near-scaling relation between [Na/H] and [Fe/H] 
([Na/H]~$\sim 0.64$~[Fe/H]) is realized despite that these 
two elements have suffered appreciable abundance changes.\\
--- (2) However, as $T_{\rm eff}$ is increased (higher than 10000~K), 
this relation first becomes appreciably loosened and ambiguous 
at the transition region (10000~K~$\ltsim T_{\rm eff} \ltsim$~11000~K),
then the correlation eventually disappears at 
11000~K~$\ltsim T_{\rm eff} \ltsim$~14000~K where [Na/H] stabilizes 
almost at the primordial value\footnote{
Considering the near-homogeneity of these 
Na abundance without significant diversity, 
we tend to regard that primordial Na abundances (when stars were formed) 
are retained in the atmospheres of these stars at $T_{\rm eff} \gtsim 11000$~K. 
As a matter of fact, we are not certain whether the slightly subsolar 
tendency in [Na/H] (by $\sim$~0.2~dex on the average) is really meaningful,
considering the possibility that our $T_{\rm eff}$ scale might as well be
systematically raised by $\sim 200$~K (cf. footnote 3).},
despite both O and Fe still show $v_{\rm e}\sin i$-dependent anomalies. 
This suggests that the physical process 
causing abundance peculiarities does not work any more on Na 
at this higher-$T{\rm eff}$ regime of late B-type stars, 
though it still operates on both Fe and O (typically exhibited by HgMn stars).\\
--- (3) It is interesting to note that the ``transition region'' stars
at 10000~K~$\ltsim T_{\rm eff} \ltsim$~11000~K are in the mass range
of 2.5--3~$M_{\odot}$ (cf. figure 1), which just correspond to the region on 
the HR diagram occupied by both HgMn stars and hot Am stars nearly overlapped
as pointed out by Adelman et al. (2003; cf. figure 1 therein).
We may state that these two groups of non-magnetic chemically peculiar 
stars (Am stars and HgMn stars) could be roughly characterized by the 
existence of Na--Fe correlation (i.e., both elements undergo abundance 
peculiarities in a similar manner) and the absence of such connection 
(i.e., Fe suffers anomaly while Na do not), respectively.

This consequence may serve as an important observational constraint 
for understanding the mechanism producing abundance peculiarities 
in upper main-sequence stars.
Generally, less attention seems to have been paid to Na in early-type
chemically peculiar stars, presumably because published observational
data have been insufficient. Above all, diffusion calculations for Na in 
the envelope of B-type stars with $M \gtsim 2.5 M_{\odot}$ seem to have 
been rarely conducted so far, except for the old work (e.g., Michaud et al. 1976).
In this respect, new contributions by theoreticians are desirably expected
toward reasonably explaining the observational trend for Na (along with O 
and Fe) confirmed in this investigation.

\bigskip

This research has made use of the SIMBAD database, operated by
CDS, Strasbourg, France.

\clearpage

\begin{figure}
  \begin{center}
    \FigureFile(100mm,100mm){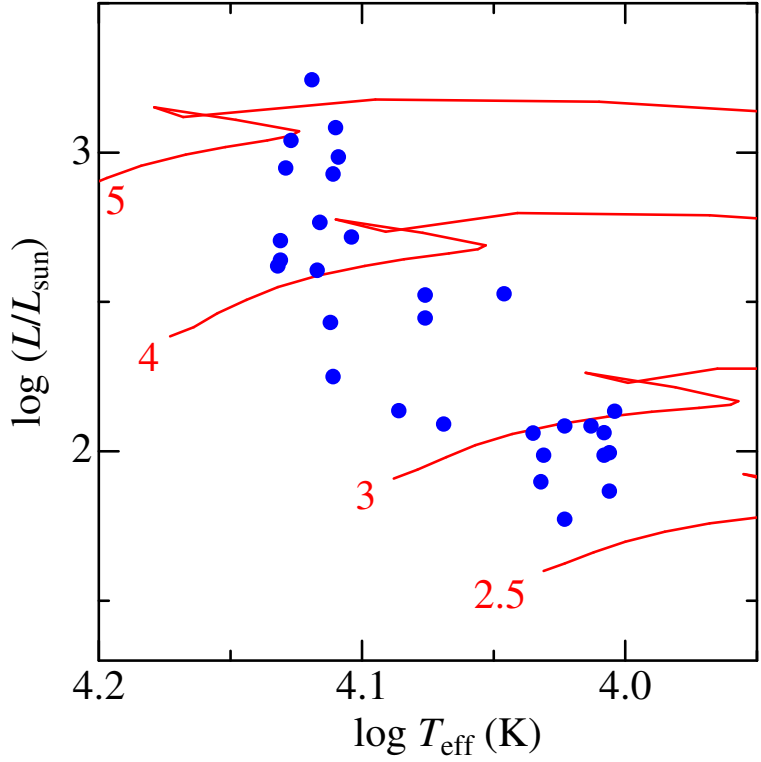}
  \end{center}
\caption{
Plots of the 30 program stars on the theoretical HR diagram
($\log (L/L_{\odot})$ vs. $\log T_{\rm eff}$), where the effective 
temperature ($T_{\rm eff}$) was determined from $uvby\beta$ photometry 
as described in section 3 and the bolometric luminosity ($L$) was 
evaluated from the apparent visual magnitude, Hipparcos parallax 
(van Leeuwen 2007), Arenou et al.'s (1992) interstellar extinction
map, and Flower's (1996) bolometric correction. 
Theoretical evolutionary tracks corresponding to the solar metallicity 
computed by Girardi et al. (2000) for four different initial masses 
(2.5, 3, 4, and 5~$M_{\odot}$) are also depicted for comparison.
}
\end{figure}

\begin{figure}
  \begin{center}
    \FigureFile(150mm,200mm){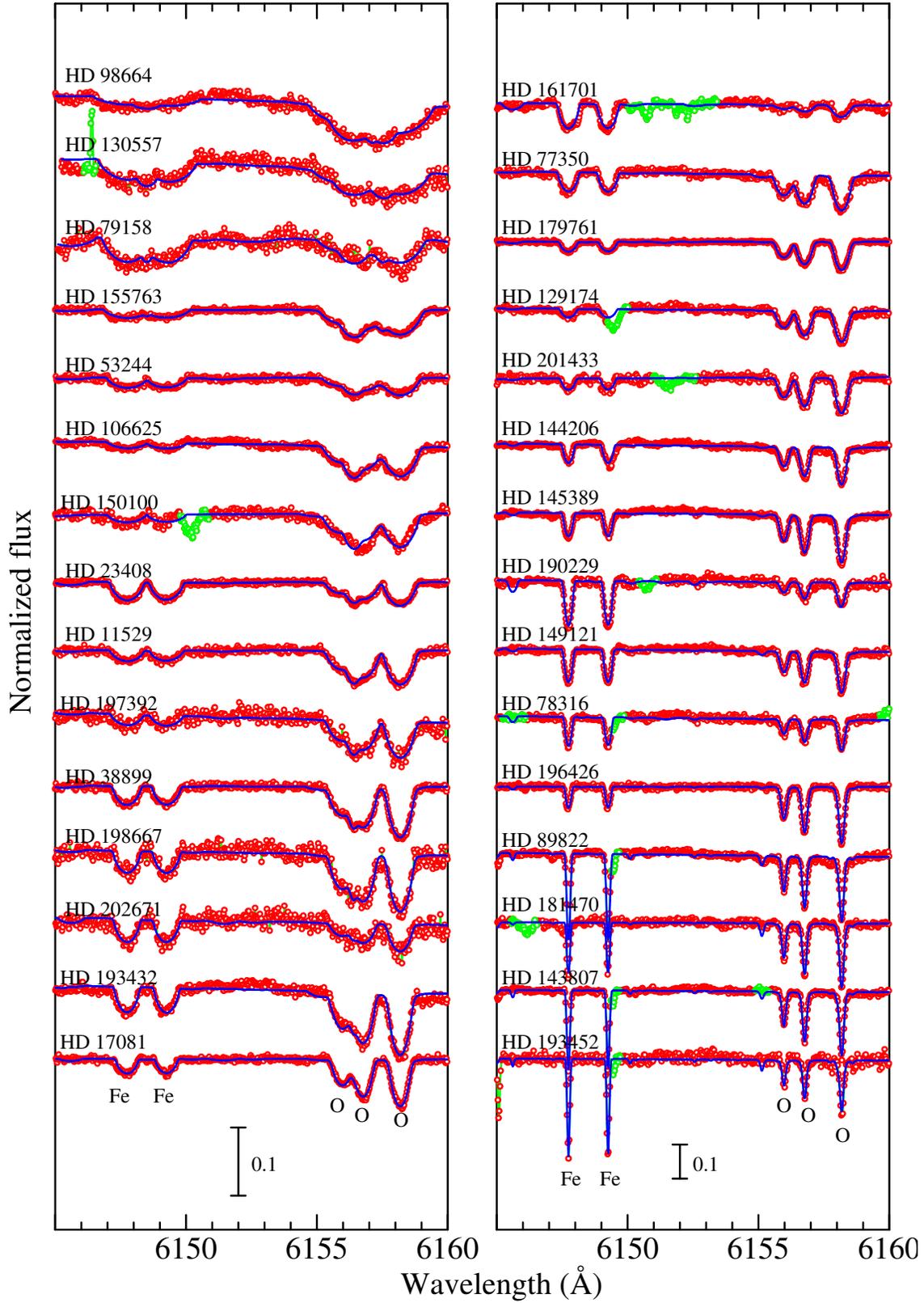}
  \end{center}
\caption{
Synthetic spectrum fitting in the 6145--6160~$\rm\AA$ region 
accomplished by varying $v_{\rm e}\sin i$, $A^{\rm O}$, and $A^{\rm Fe}$.
The best-fit theoretical spectra are shown by blue solid lines. 
The observed data are plotted by symbols, where those due to
stellar origin and actually used in fitting are depicted in red,
while those of non-stellar origin and rejected in fitting are
highlighted in green.The spectra are arranged (from top to bottom) 
in the descending order of $v_{\rm e}\sin i$ as in table 1, 
and offset values of 0.1 (left panel ) and 0.2 (right panel) are 
applied to each spectrum relative to the adjacent one. 
}
\end{figure}

\begin{figure}
  \begin{center}
    \FigureFile(150mm,200mm){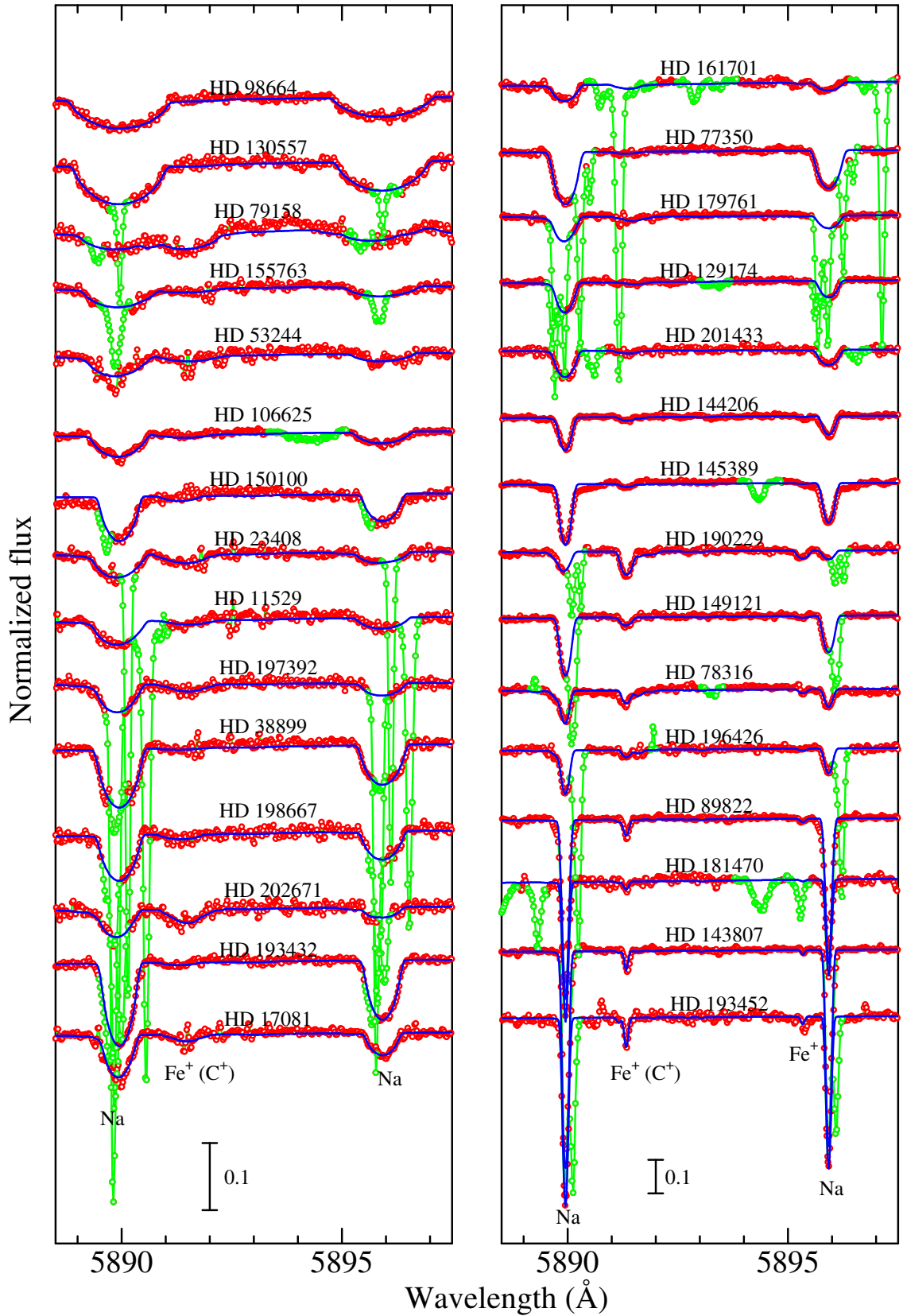}
  \end{center}
\caption{
Synthetic spectrum fitting in the 5888.5--5897.5~$\rm\AA$ region 
accomplished by varying $A^{\rm Na}$ and $A^{\rm Fe}$.
Note that this spectral region tends to be severely contaminated by sharp
interstellar Na~{\sc i} lines, which had to be masked in the fitting.
Otherwise, the same as in figure 2.
}
\end{figure}

\begin{figure}
  \begin{center}
    \FigureFile(120mm,160mm){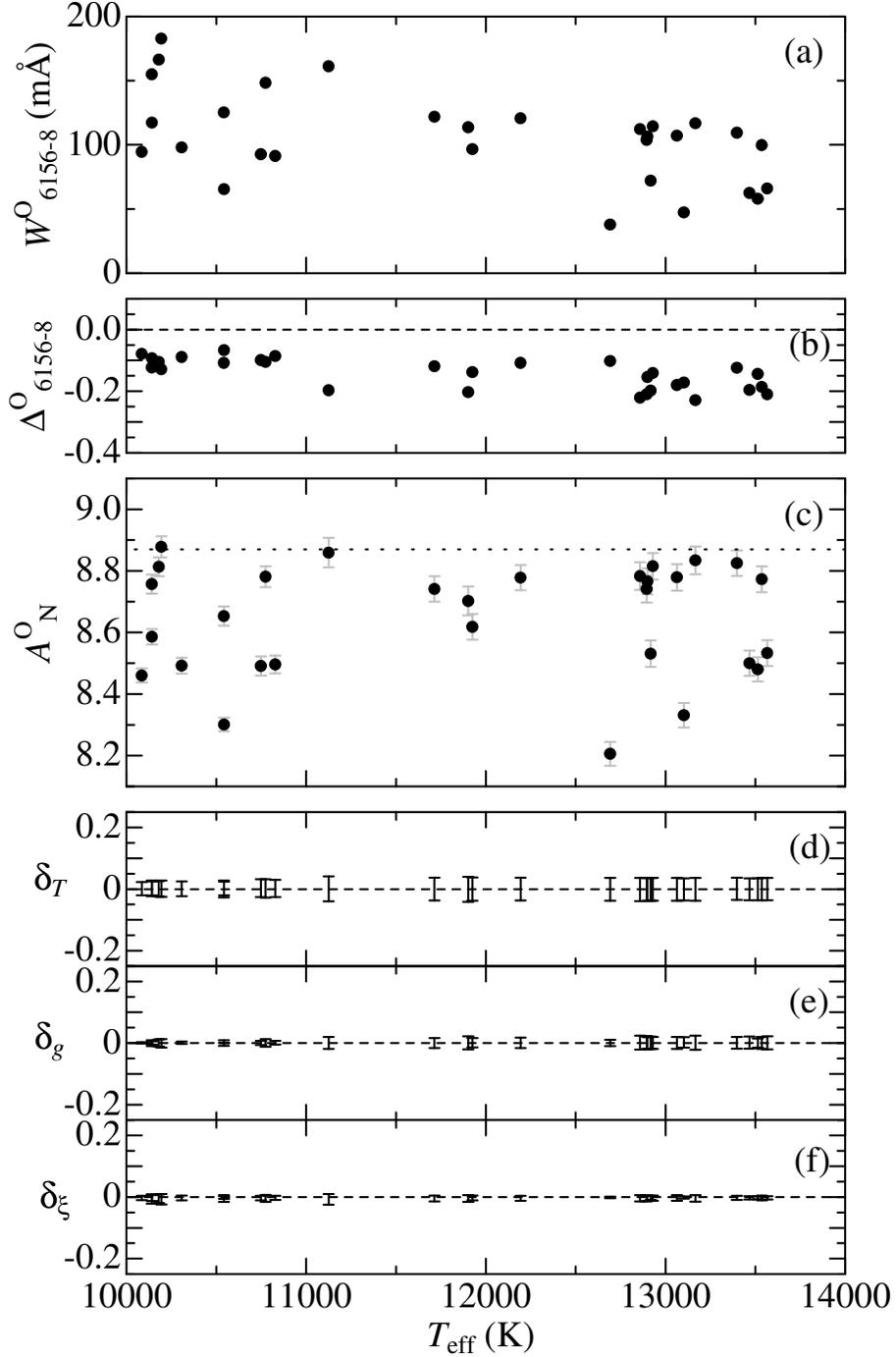}
  \end{center}
\caption{
Oxygen abundances along with the abundance-related quantities 
specific to the O~{\sc i} 6156--6158 triplet feature, plotted against $T_{\rm eff}$. 
(a) $W^{\rm O}_{6156-8}$ (equivalent width), 
(b) $\Delta^{\rm O}_{6156-8}$ (non-LTE correction),
(c) $A^{\rm O}_{\rm N}$ (non-LTE oxygen abundance, where the attached error bar
is $\delta_{Tg\xi}$ defined in subsection 4.2), 
(d) $\delta_{T+}$ and $\delta_{T-}$ (abundance variations 
in response to $T_{\rm eff}$ changes of +3\% and $-3$\%), 
(e) $\delta_{g+}$ and $\delta_{g-}$ (abundance variations 
in response to $\log g$ changes of $+0.2$~dex and $-0.2$~dex), 
and (f) $\delta_{\xi +}$ and $\delta_{\xi -}$ (abundance 
variations in response to perturbing the standard $\xi = 1$~km~s$^{-1}$
by +1~km~s$^{-1}$ and $-1$~km~s$^{-1}$).
The signs of $\delta$'s are $\delta_{T+}>0$, $\delta_{T-}<0$,
$\delta_{g+}<0$, $\delta_{g-}>0$, $\delta_{\xi +}<0$, and
$\delta_{\xi -}>0$. The horizontal dotted line in panel (c) indicates
the reference solar abundance (cf. the caption in figure 7).
}
\end{figure}

\begin{figure}
  \begin{center}
    \FigureFile(120mm,160mm){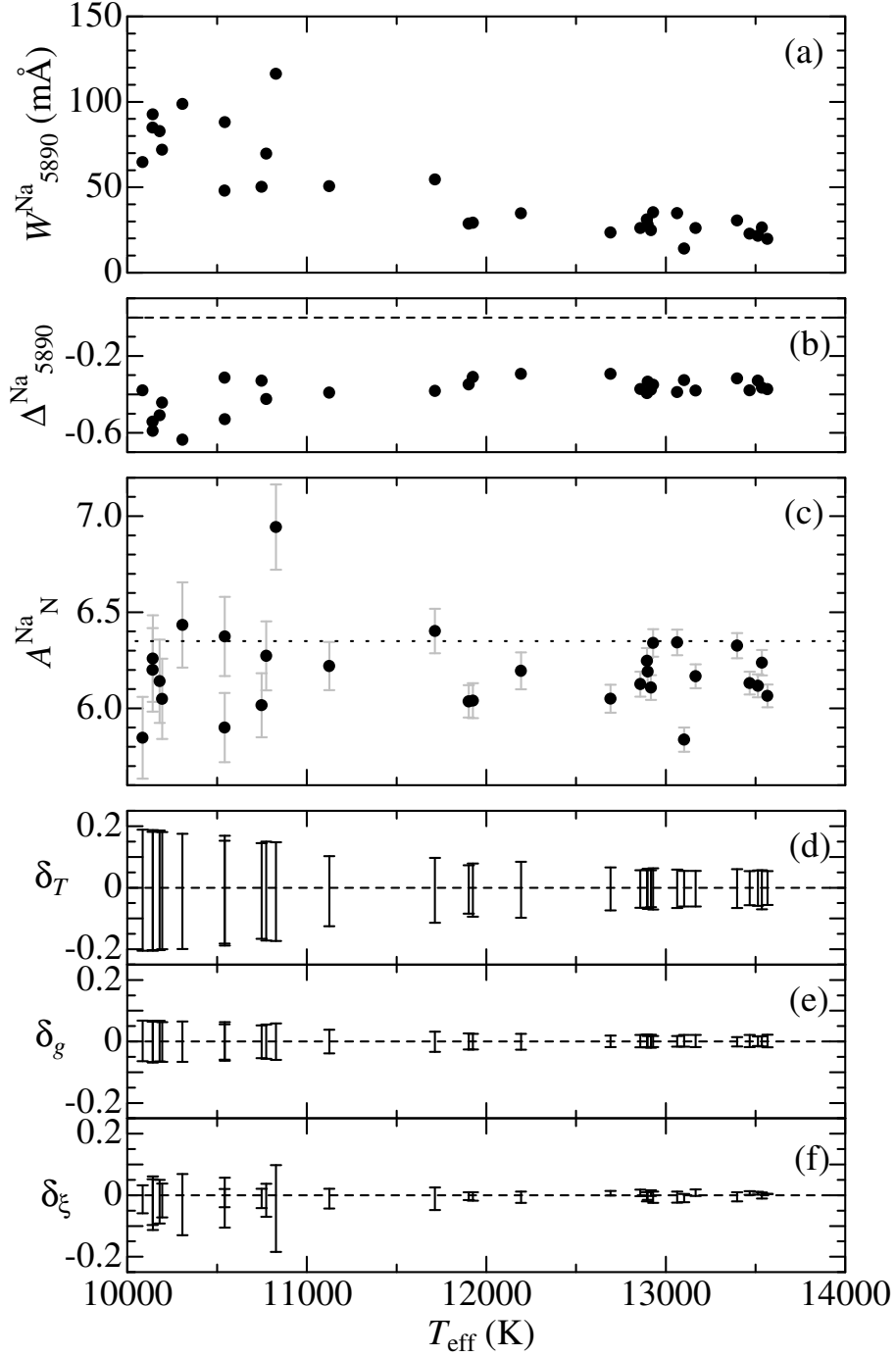}
  \end{center}
\caption{
Sodium abundances along with the abundance-related quantities 
specific to the Na~{\sc i} 5890 line, plotted against $T_{\rm eff}$. 
(a) $W^{\rm Na}_{5890}$ (equivalent width), 
(b) $\Delta^{\rm Na}_{5890}$ (non-LTE correction),
(c) $A^{\rm Na}_{\rm N}$ (non-LTE sodium abundance), 
(d) $\delta_{T+}$ and $\delta_{T-}$ (abundance variations 
in response to changing $T_{\rm eff}$), 
(e) $\delta_{g+}$ and $\delta_{g-}$ (abundance variations 
in response to changing $\log g$ changes), 
and (f) $\delta_{\xi +}$ and $\delta_{\xi -}$ (abundance 
variations in response to changing $\xi$).
The signs of $\delta$'s are $\delta_{T+}>0$, $\delta_{T-}<0$,
$\delta_{g+}<0$, $\delta_{g-}>0$, $\delta_{\xi +}<0$, and
$\delta_{\xi -}>0$.
Otherwise, the same as figure 4.}
\end{figure}

\begin{figure}
  \begin{center}
    \FigureFile(120mm,160mm){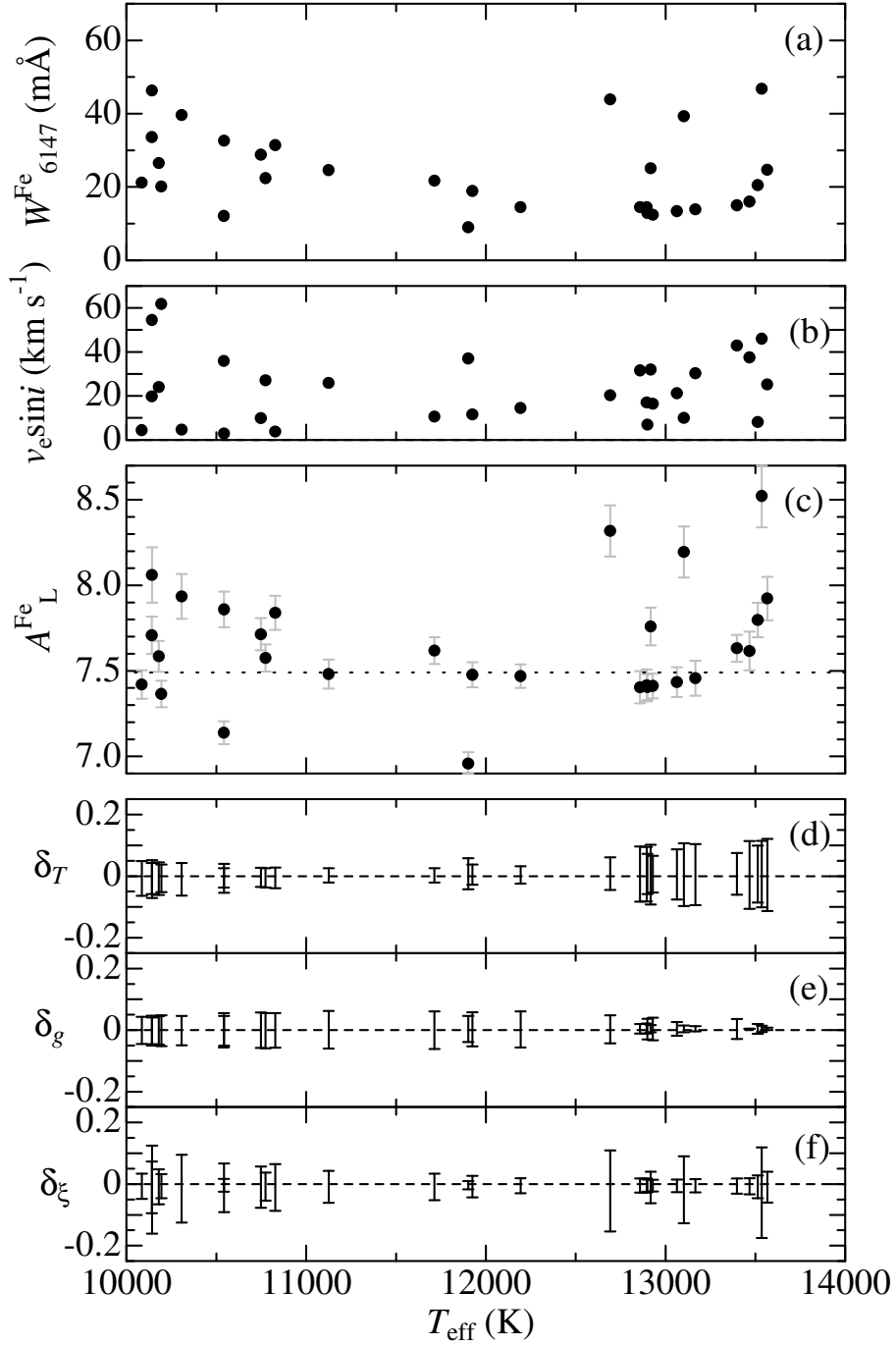}
  \end{center}
\caption{
Iron abundances along with the abundance-related quantities 
specific to the Fe~{\sc ii} 6147 line, plotted against $T_{\rm eff}$. 
(a) $W^{\rm Fe}_{6147}$ (equivalent width), 
(b) $v_{\rm e}\sin i$ (projected rotational velocity),
(c) $A^{\rm Fe}_{\rm L}$ (LTE iron abundance), 
(d) $\delta_{T+}$ and $\delta_{T-}$ (abundance variations 
in response to changing $T_{\rm eff}$), 
(e) $\delta_{g+}$ and $\delta_{g-}$ (abundance variations 
in response to changing $\log g$ changes), 
and (f) $\delta_{\xi +}$ and $\delta_{\xi -}$ (abundance 
variations in response to changing $\xi$).
The signs of $\delta$'s are $\delta_{T+}>0$, $\delta_{T-}<0$,
$\delta_{g+}>0$, $\delta_{g-}<0$, $\delta_{\xi +}<0$, and
$\delta_{\xi -}>0$.
Otherwise, the same as figure 4.}
\end{figure}

\begin{figure}
  \begin{center}
    \FigureFile(130mm,170mm){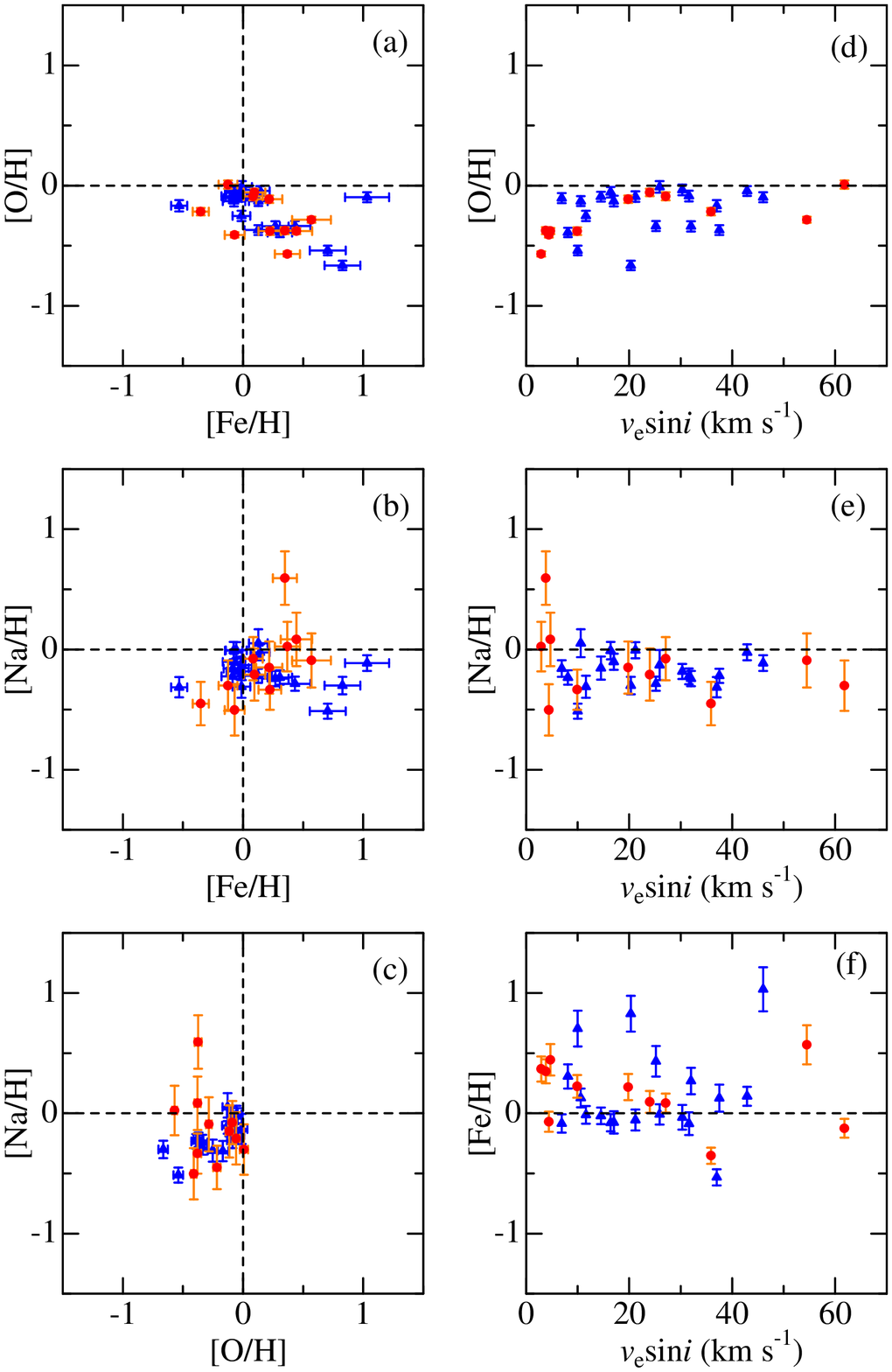}
  \end{center}
\caption{
Left panels show the correlations between [O/H], [Na/H], and [Fe/H]: 
(a) [O/H] vs. [Fe/H], (b) [Na/H] vs. [Fe/H], and (c) [Na/H] vs. [O/H].
In right panels (d), (e), and (f),  [O/H], [Na/H], and [Fe/H] are plotted 
against $v_{\rm e} \sin i$, respectively.
Circles (red) and triangles (blue) correspond the results for stars with 
$T_{\rm eff} < 11000$~K and those with $T_{\rm eff} > 11000$~K, respectively.
In evaluating [X/H] ($\equiv A^{\rm X}_{\rm star} - A^{\rm X}_{\rm sun}$),
we adopted $A^{\rm O}_{\rm sun} = 8.87$, $A^{\rm Na}_{\rm sun} = 6.35$, 
and $A^{\rm Fe}_{\rm sun} = 7.49$, for which the abundances of Procyon
(known to be almost of solar composition) derived from the same lines 
were substituted (cf. Takeda et al. 2008, Paper~I).}
\end{figure}

\begin{figure}
  \begin{center}
    \FigureFile(100mm,100mm){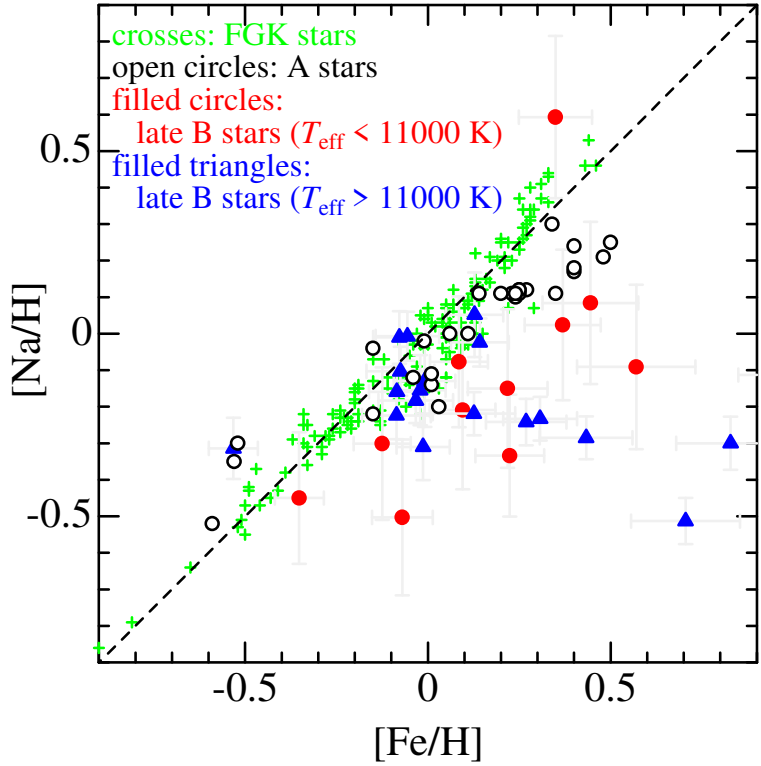}
  \end{center}
\caption{
Combined [Na/H] vs. [Fe/H] correlations for stars of different spectral types.
Crosses $\cdots$ FGK stars (Takeda 2007), open circles $\cdots$ sharp-lined
A-type stars (Paper~II), filled circles $\cdots$ late B-type stars
with $T_{\rm eff} < 11000$~K (this study), and filled triangles $\cdots$
late B-type stars with $T_{\rm eff} > 11000$~K (this study).
}
\end{figure}

\setcounter{table}{0}
\begin{table}[h]
\scriptsize
\caption{Basic data of the program stars and the results of the analysis.}
\begin{center}
\begin{tabular}
{ccc ccc ccc ccc cc l}\hline \hline
HD\# & Desig. & Sp.type & $T_{\rm eff}$ & $\log g$ & $v_{\rm e} \sin i$ &
$W_{6156-8}^{\rm O}$ & $A_{\rm N}^{\rm O}$ & $\Delta_{6156-8}^{\rm O}$ &
$W_{5890}^{\rm Na}$ & $A_{\rm N}^{\rm Na}$ & $\Delta_{5890}^{\rm Na}$ &
$W_{6147}^{\rm Fe}$ & $A_{\rm L}^{\rm Fe}$ & Obs. date \\
\hline
098664 & $\sigma$ Leo & B9.5Vs & 10194 & 3.75 & 62 & 183 & 8.88 & $-$0.13 & 72 & 6.05 & $-$0.44 & 20 & 7.37 & 2012 May 31\\
130557 & HR 5522 & B9Vsvar... & 10142 & 3.85 & 55 & 117 & 8.59 & $-$0.09 & 93 & 6.26 & $-$0.59 & 46 & 8.06 & 2012 May 31\\
079158 & 36 Lyn & B8IIIMNp & 13535 & 3.72 & 46 & 100 & 8.77 & $-$0.19 & 26 & 6.24 & $-$0.37 & 47 & 8.52 & 2012 May 31\\
155763 & $\zeta$ Dra & B6III & 13397 & 4.24 & 43 & 109 & 8.83 & $-$0.12 & 31 & 6.33 & $-$0.32 & 15 & 7.63 & 2006 Oct 23\\
053244 & $\gamma$ CMa & B8II & 13467 & 3.42 & 38 & 63 & 8.50 & $-$0.20 & 23 & 6.13 & $-$0.38 & 16 & 7.62 & 2006 Oct 23\\
106625 & $\gamma$ Crv & B8III & 11902 & 3.36 & 37 & 114 & 8.70 & $-$0.20 & 29 & 6.04 & $-$0.35 & 9 & 6.96 & 2012 May 27\\
150100 & 16 Dra & B9.5Vn & 10542 & 3.84 & 36 & 125 & 8.65 & $-$0.11 & 48 & 5.90 & $-$0.31 & 12 & 7.14 & 2012 May 31\\
023408 & 20 Tau & B8III & 12917 & 3.36 & 32 & 72 & 8.53 & $-$0.20 & 25 & 6.11 & $-$0.38 & 25 & 7.76 & 2006 Oct 23\\
011529 & 46 Cas & B8III & 12858 & 3.43 & 32 & 112 & 8.78 & $-$0.22 & 26 & 6.13 & $-$0.37 & 15 & 7.40 & 2006 Oct 23\\
197392 & HR 7926 & B8II-III & 13166 & 3.46 & 30 & 117 & 8.83 & $-$0.23 & 26 & 6.17 & $-$0.38 & 14 & 7.46 & 2012 May 29\\
038899 & 134 Tau & B9IV & 10774 & 4.02 & 27 & 148 & 8.78 & $-$0.11 & 70 & 6.27 & $-$0.42 & 22 & 7.58 & 2006 Oct 23\\
198667 & 5 Aqr & B9III & 11125 & 3.42 & 26 & 161 & 8.86 & $-$0.20 & 51 & 6.22 & $-$0.39 & 25 & 7.48 & 2012 May 28\\
202671 & 30 Cap & B5II/III & 13566 & 3.36 & 25 & 66 & 8.53 & $-$0.21 & 20 & 6.07 & $-$0.37 & 25 & 7.92 & 2012 May 31\\
193432 & $\nu$ Cap & B9IV & 10180 & 3.91 & 24 & 166 & 8.81 & $-$0.11 & 83 & 6.14 & $-$0.51 & 27 & 7.59 & 2012 May 26\\
017081 & $\pi$ Cet & B7IV & 13063 & 3.72 & 21 & 107 & 8.78 & $-$0.18 & 35 & 6.34 & $-$0.39 & 13 & 7.43 & 2006 Oct 20\\
161701 & HR 6620 & B9V & 12692 & 4.04 & 20 & 38 & 8.21 & $-$0.10 & 24 & 6.05 & $-$0.29 & 44 & 8.32 & 2012 May 27\\
077350 & $\nu$ Cnc & A0III & 10141 & 3.68 & 20 & 155 & 8.76 & $-$0.12 & 85 & 6.20 & $-$0.54 & 34 & 7.71 & 2012 May 27\\
179761 & 21 Aql & B8II-III & 12895 & 3.46 & 17 & 104 & 8.74 & $-$0.21 & 31 & 6.25 & $-$0.39 & 15 & 7.42 & 2006 Oct 21\\
129174 & $\pi^1$ Boo & B9p MnHg & 12929 & 4.02 & 16 & 114 & 8.82 & $-$0.14 & 35 & 6.34 & $-$0.35 & 12 & 7.41 & 2012 May 26\\
201433 & HR 8094 & B9V & 12193 & 4.24 & 15 & 121 & 8.78 & $-$0.11 & 35 & 6.20 & $-$0.29 & 15 & 7.47 & 2012 May 28\\
144206 & $\upsilon$ Her & B9III & 11925 & 3.79 & 12 & 97 & 8.62 & $-$0.14 & 29 & 6.04 & $-$0.31 & 19 & 7.48 & 2012 May 27\\
145389 & $\phi$ Her & B9MNp... & 11714 & 4.02 & 11 & 122 & 8.74 & $-$0.12 & 55 & 6.40 & $-$0.38 & 22 & 7.62 & 2012 May 27\\
190229 & 14 Sge & B9MNp... & 13102 & 3.46 & 10 & 47 & 8.33 & $-$0.17 & 14 & 5.84 & $-$0.33 & 39 & 8.20 & 2012 May 28\\
149121 & 28 Her & B9.5III & 10748 & 3.89 & 10 & 93 & 8.49 & $-$0.10 & 50 & 6.02 & $-$0.33 & 29 & 7.71 & 2012 May 27\\
078316 & $\kappa$ Cnc & B8IIIMNp & 13513 & 3.85 & 8 & 58 & 8.48 & $-$0.14 & 22 & 6.12 & $-$0.33 & 21 & 7.80 & 2012 May 28\\
196426 & HR 7878 & B8IIIp & 12899 & 3.89 & 7 & 106 & 8.77 & $-$0.15 & 28 & 6.19 & $-$0.33 & 13 & 7.41 & 2006 Oct 21\\
089822 & HR 4072 & A0sp... & 10307 & 3.89 & 5 & 98 & 8.49 & $-$0.09 & 99 & 6.43 & $-$0.64 & 40 & 7.94 & 2012 May 27\\
181470 & HR 7338 & A0III & 10085 & 3.92 & 4 & 95 & 8.46 & $-$0.08 & 65 & 5.85 & $-$0.38 & 21 & 7.42 & 2006 Oct 21\\
143807 & $\iota$ CrB & A0p... & 10828 & 4.06 & 4 & 91 & 8.50 & $-$0.09 & 116 & 6.94 & $-$0.71 & 31 & 7.84 & 2012 May 26\\
193452 & HR 7775 & B9.5III/IV & 10543 & 4.15 & 3 & 65 & 8.30 & $-$0.07 & 88 & 6.37 & $-$0.53 & 33 & 7.86 & 2012 May 29\\
\hline
\end{tabular}
\end{center}
Note. \\
In columns 1 through 6 are given the HD number, star designation, 
spectral type (taken from Hipparcos catalogue), effective 
temperature (in K), logarithmic surface gravity (in cm~s$^{-2}$),
and  projected rotational velocity (in km~s$^{-1}$; from 6145--6160~$\rm\AA$
fitting). Columns 7 through 14 show the results of abundance analyses:
merged equivalent width of the O~{\sc i}~6156--8 lines, non-LTE O abundance  
and NLTE correction; equivalent width of the Na~{\sc i} 5890 line, 
non-LTE Na abundance, and NLTE correction; equivalent width of the 
Fe~{\sc ii} 6147 line and (LTE) Fe abundance. Finally, column 15 gives 
the observation date. All abundance results are expressed in the usual 
normalization of $A^{\rm H}=12.00$.
The objects are arranged in the descending order of $v_{\rm e} \sin i$. 
\end{table}

\setcounter{table}{1}
\begin{table}[h]
\small
\caption{Atomic data of important lines relevant to the analysis.}
\begin{center}
\begin{tabular}
{ccclrl}\hline \hline
Desig. & Species & RMT & $\lambda (\rm\AA) $ & $\chi_{\rm low}$~(eV)  & $\log gf$ \\
\hline
6156   & O~{\sc i} & 10 &  6155.961 & 10.74 & $-1.40$ \\
       & O~{\sc i} & 10 &  6155.971 & 10.74 & $-1.05$ \\
       & O~{\sc i} & 10 &  6155.989 & 10.74 & $-1.16$ \\
6157   & O~{\sc i} & 10 &  6156.737 & 10.74 & $-1.52$ \\
       & O~{\sc i} & 10 &  6156.755 & 10.74 & $-0.93$ \\
       & O~{\sc i} & 10 &  6156.778 & 10.74 & $-0.73$ \\
6158   & O~{\sc i} & 10 &  6158.149 & 10.74 & $-1.89$ \\
       & O~{\sc i} & 10 &  6158.172 & 10.74 & $-1.03$ \\
       & O~{\sc i} & 10 &  6158.187 & 10.74 & $-0.44$ \\
\hline
5890   & Na~{\sc i} & 1 &  5889.951  & 0.00 & $+0.12$ \\
5896   & Na~{\sc i} & 1 &  5895.924  & 0.00 & $-0.18$ \\
\hline
6147   & Fe~{\sc ii} & 74 &  6147.741 & 3.89 & $-2.72$ \\
6149   & Fe~{\sc ii} & 74 &  6149.258 & 3.89 & $-2.72$ \\
\hline
\end{tabular}
\end{center}
\scriptsize
Note. \\
All data are were taken from Kurucz \& Bell's (1995) compilation.
RMT is the multiplet number in the Revised Multiplet Table (Moore 1959).
\end{table}

\setcounter{table}{2}
\begin{table}[h]
\scriptsize
\caption{Statistical trend of correlation between Na and Fe for each group of different $T_{\rm eff}$.}
\begin{center}
\begin{tabular}
{ccccccccl}\hline \hline
Source & Sp. type & $T_{\rm eff}$ range & $N$ & $r$ & $m$ & $b$ & $\sigma$ & Remark\\
\hline
Takeda (2007) & FGK & 5000--7000~K & 159 & +0.962 & $+1.031(\pm 0.023)$ & $-0.016(\pm 0.006)$ & 0.078 & \\
Paper~II & A & 7000--10000~K & 28 & +0.945 & $+0.641(\pm 0.043)$ & $-0.056(\pm 0.013)$ & 0.065 & \\
This study & late B (near A0) & 10000--11000~K & 10 & +0.782 & $+0.546(\pm 0.154)$ & $-0.280(\pm 0.047)$ & 0.122 & HD~143807 (outlier) excluded\\
This study & late B & 11000--14000~K & 19 & $-0.267$ & $-0.097(\pm 0.085)$ & $-0.167(\pm 0.034)$ & 0.131 & \\
  & {\it late B (subdivided)} &  {\it 11000--13000~K} & $\mathit{11}$ & $\mathit{-0.085}$ & $\mathit{-0.032(\pm 0.123)}$ & $\mathit{-0.171(\pm 0.039)}$ & $\mathit{0.121}$ & \\ 
  & {\it late B (subdivided)} &  {\it 13000--14000~K} & $\mathit{8}$ & $\mathit{-0.407}$ & $\mathit{-0.174(\pm 0.159)}$ & $\mathit{-0.139(\pm 0.077)}$ & $\mathit{0.148}$ & \\ 
\hline
\end{tabular}
\end{center}
\scriptsize
Note. \\
See figure 8 for graphical representations of Na vs. Fe correlations for each group. 
The last two lines (printed in {\it italic} type) show the results for the two 
subdivided groups of the $T_{\rm eff}$ = 11000--14000~K stars given in the 4th line. 
$N$ is the number of stars, and $r$ is the correlation coefficient
between [Na/H] and [Fe/H]. $m$ and $b$ are the coefficients of the
linear regression line ([Na/H] = $m$~[Fe/H] + $b$) determine by the least squares
analysis (standard errors of $m$ and $b$ are also given in parentheses with $\pm$), 
while $\sigma$ is the standard deviation of the residual between the actual 
values of [Na/H] and those computed from this linear relation.
\end{table}

\end{document}